\pdfoutput=1 

\documentclass[12pt]{article}

\usepackage{amssymb}
\usepackage{caption} 
\usepackage{booktabs} 
\usepackage{multirow}

\usepackage{moreverb,url}

\usepackage{hyperref} 
\hypersetup{colorlinks,linkcolor=blue,bookmarksopen,bookmarksnumbered,citecolor=blue,urlcolor=blue} 

\usepackage{enumitem} 

 \usepackage[numbers]{natbib} 
 
\usepackage{wasysym} 

\usepackage{threeparttable} 
\usepackage{longtable} 

\begin{document}

\title{The Good, The Bad and The Missing: A Narrative Review of Cyber-security Implications for Australian Small Businesses}

\author{Tracy Tam, Asha Rao and Joanne Hall}

\date{RMIT University,\\
124 Latrobe St,\\
Melbourne 3000,\\
Australia.} 

\maketitle

\begin{abstract}
Small businesses (0-19 employees) are becoming attractive targets for cyber-criminals, but struggle to implement cyber-security measures that large businesses routinely deploy. There is an urgent need for effective and suitable cyber-security solutions for small businesses as they employ a significant proportion of the workforce. 

In this paper, we consider the small business cyber-security challenges not currently addressed by research or products, contextualised via an Australian lens.  We also highlight some unique characteristics of small businesses conducive to cyber-security actions.

Small business cyber-security discussions to date have been narrow in focus and lack re-usability beyond specific circumstances. Our study uses global evidence from industry, government and research communities across multiple disciplines. We explore the technical and non-technical factors negatively impacting a small business' ability to safeguard itself, such as resource constraints, organisational process maturity, and legal structures.  Our research shows that some small business characteristics, such as agility, large cohort size, and piecemeal IT architecture, could allow for increased cyber-security.  

We conclude that there is a gap in current research in small business cyber-security. In addition, legal and policy work are needed to help small businesses become cyber-resilient. 

\end{abstract}


\section{Introduction}
Businesses around the world are under pressure to use and adopt technology, with the COVID-19 pandemic accelerating this need \cite{ABCNews2020}.  At the same time, cyber-crimes are becoming increasingly sophisticated \cite{Symantec2019} with the Australian Government \cite{AustralianGovernment2016} estimating the cost of cyber-crime to the Australian economy at over AU\$15 billion annually. Small businesses employ nearly half of the Australian private sector workforce \cite{DepartmentofParliamentaryServices2018} and need to be supported against cyber-criminals.  

Anecdotes \cite{Armstrong2019, Atwood2019,Bainbridge2020}, industry \cite{hiscox2019, Symantec2019} and government \cite{AustralianCyberSecurityCentre2020a} studies indicate that small businesses and individuals are increasingly becoming cyber-crime targets.  A small business can lose data and customers, and incurs compliance penalties from a cyber-security incident.  These penalties include fines greater of 4\% of global turnover or 20 million for non-compliance to the General Data Protection Regulation (GDPR) \cite{CouncilofTheEuropeanUnion2016} in the European Union, and levies up to US\$1.8 million per violation \cite{Azar2020} of HIPAA \cite{CDC2018} in the US. In Australia, most small businesses with less than AU\$3 million turnover are exempt from the Privacy Act \cite{OfficeofParliar2019}, but this exemption may not remain in the future \cite{Kelkar2013, Newbury2017}.  Beyond these penalties, the importance and practicalities of small businesses preparing for and recovering from cyber-incidents are still poorly articulated and communicated. 

Previous research has considered the Australian small and medium enterprise (SME) cyber-security situation from different perspectives. Valli et al \cite{Valli2014} survey Western Australian SMEs and call for cyber-security education for SMEs, while Dojkovki et al. \cite{Dojkovski2010,Dojkovski2007, Dojkovski2006} highlight the influence of Australian culture on SME cyber-security.  Internationally, reviews of SME cyber-security (albeit with different size definitions) emphasise ineffective security management practices \cite{Alahmari2020} and differences between SMEs and large corporations \cite{Heidt2019,Osborn2017}. Other research offers global insights into culture \cite{Ruhwanya2019}, strategies \cite{Bryan2020}, policies \cite{Almeida2018} and specific technologies \cite{Salleh2018}.  Several surveys on cyber-security, for example \cite{AustralianCyberSecurityCentre2020a, CynchSecurity2021, NSWGovernment2017}, focus on self-reported data on cyber-security from Australian SMEs.  

This paper uses existing global research as well as evidence from multiple disciplines  to identify gaps and opportunities for Australian small business cyber-security.  We look in detail at the characteristics of Australian small businesses that may be impeding their cyber-security posture.  We find that certain characteristics could help small businesses better secure themselves.

The definition we use here, of 0-19 employees, covers a significant proportion of business communities around the world. In 2015, around 29 million US firms employed 0-19 employees \cite{Luque2019,USCensus2015}, while in 2019, in the UK over 98\% of private sector businesses employed 0-49 people \cite{BritishCrown2019}. Thus, this research has relevance to small businesses from other countries with similar societal profiles to Australia \cite{Hofstede2001,OECD2018a}, since human owners run small businesses \cite{Moran1998,Kickul2002}.

This paper is structured as follows: In the \hyperref[background]{next} section we examine the background to business cyber-security and the Australian small business context. Section~\ref{review} looks at existing review articles and compares them to our study. Section~\ref{research} discusses the difficulties in research and data collection for small business cyber-security. Section~\ref{constraints} looks at the constraints faced by small businesses in implementing cyber-security, while section~\ref{opportunities} highlights potential opportunities that unique small business characteristics present with respect to cyber-security. We end with a conclusion and suggestions for future work.

\section{Background to Small Business Cyber-security}
\label{background}

Small business cyber-security has some commonalities with large business cyber-security. The traditional aim of cyber-security is to protect business IT infrastructure, and the information essential for the business' day to day operations.  However evidence has since emerged that the benefit of business cyber-security extends beyond this, to its people, processes and assets. 

The cyber-security industry is currently worth over US\$100 billion globally \cite{IDC2019}.  While cyber-security previously focused on maintaining the confidentiality, integrity and availability \cite{Kaur2013} of information systems and data, current best practice includes non-technical aspects such as the people using the information (security culture) \cite{Cheng2013,Sawaya2017}. A one-size-fits-all cyber-security solution does not exist \cite{VonSolms2004,DArcy2009,Bernik2016}. Large businesses have adopted a variety of measures to secure their information and IT assets \cite{Deloitte2019} against cyber-security threats.  

The benefits of securing a business extends beyond protecting the assets owned directly by the business.  The disruptions to individuals from data breaches such as Equifax \cite{115thCongress2018} and Ticketmaster \cite{Chau2018} highlight the societal cost.  The financial cost of an incident can severely disrupt the personal lives of small business owners, putting livelihoods in jeopardy \cite{Bainbridge2020a}. 
The high level of codependency between people, technologies and processes has resulted in modern security standards/frameworks including ``processes, organisational structures, policies and procedures, information flows, culture and behaviors, skills and infrastructure'' \cite[P.13]{ISACA2019c}.

\subsection{The Australian Small Business Context}
In Australia, over 40\% of private sector employment is from micro (0-4 employees) and small (5-19 employees) businesses \cite{DepartmentofParliamentaryServices2018}.  The Australian Bureau of Statistics (ABS) defines a small business as one with 0-19 employees \cite{AustralianBureauofStatistics8167}.  In 2016, 42\% of Australian small businesses believed limiting online presence protected their business \cite{NSWGovernment2017}. 
In the 2018/19 financial year, only 40.3\% of small businesses received orders via the Internet\footnote{Note: This includes any sale of services or goods that have been committed to/ordered via the Internet.} \cite{AustralianBureauofStatistics2020}.

Technology avoidance is no longer a feasible strategy for managing cyber-security risks. Online sales are becoming more attractive, influenced by market forces and official bodies \cite{AustralianProductivityCommissionandNewZealandProductivityCommission2019} with the social distancing requirements of COVID-19 only hastening this 
\cite{ABCNews2020,StateGovernmentofVictoria2020a}.  

With nearly half of Australian small businesses allocating less than AU\$500 annually to cyber-security \cite{AustralianCyberSecurityCentre2020a}, they cannot afford cyber-security expertise.  Without external help, small businesses cannot secure themselves, and not just in times of emergency \cite{Woschke2017}.

\section{Existing Reviews}
\label{review}

There is considerable literature, both academic and industry, on small and medium business (or SME) cyber-security.  Here we identify the literature examining cyber-security challenges. 

\subsection{Search Method}
A keyword search was conducted within 5 academic search engines, namely Scopus, ProQuest, Science Direct, SpringerLink and IEEE Xplore. The keywords used were kept broad to maximise the number of results returned and constructed from the core concepts of ``small business'', ``cyber-security'' and ``literature review''. These are listed below:
\begin{enumerate}
\item Small business - ``small business'', ``small enterprise'' or ``SME''.
\item Cyber security - Various spellings of ``cyber-security'', ``IT security'' or ``information security''.
\item Literature review - ``review'' or ``survey'' with ``literature''
\end{enumerate}
The keyword pertaining to ``literature review'' (3) was replaced with a search filter of review articles where this option was offered by the database. Where possible, the search was limited to title, abstract and keywords.  To ensure relevance to today's context, the search was restricted to publications dated on or after 2016. No filtering was done based on journal or conference ranking.

Due to the general nature of the keywords, the initial searches returned a total of 2759 papers.  Each of these papers was examined to determine whether the primary focus was small business (or SME) cyber-security.  Most importantly, we looked for papers that exclusively discussed their cyber-security topic in a small business/SME context, rather than treating the cohort as a sub-discussion. Based on this criteria 22 papers were shortlisted.  A detailed examination of these papers gave us one very relevant review paper \cite{Alahmari2020}, and three partially relevant papers \cite{Heidt2019,Suryotrisongko2019,Salleh2018}.

\subsection{Comparing Research Aims}
Table~\ref{reviewfocus} compares the aims of our paper to the selected review papers,  based on the following contexts:\\
\begin{minipage}{\linewidth}
\begin{enumerate}[label=\Alph*]
\item Restricts context to small business.
\label{criteria}
\item Critical examines  landscape of existing small business cyber-security research.
\item Examines potential technical issues with small business use of cyber-security solutions. 
\item Examines potential human issues for small business.
\item Identifies structural opportunities for small businesses.
\end{enumerate}
\end{minipage}

\begin{table}[h!]
\small\sf\centering
\caption{Existing Literature, Reviews and Focus Comparison on Small Business Cyber-security. (\checked  denotes areas addressed.  \textasciitilde  denotes areas partially addressed. Blank denotes areas not addressed.) \\See page~\pageref{criteria} for criteria used for comparison. \label{reviewfocus}}
\begin{tabular}{p{2.25cm}p{1.75cm}p{2.5cm}p{0.75cm}p{0.75cm}p{0.75cm}p{0.75cm}p{0.75cm}}
\toprule
Paper&Literature Type&Focus&A&B&C&D&E\\
\midrule
\rule{0pt}{4ex}Alahmari \& Duncan \cite{Alahmari2020}&Academic&Review&&&&\checked&\textasciitilde\\
\midrule
\rule{0pt}{4ex}Heidt et al. \cite{Heidt2019}&Academic&IT Security Investment&&\textasciitilde&&&\\
\rule{0pt}{4ex}Suryotrisongko \& Musashi \cite{Suryotrisongko2019}&Academic&Cyber-security Research Taxonomy Review&&\textasciitilde&&\textasciitilde&\\
\rule{0pt}{4ex}Salleh et al. \cite{Salleh2018}&Academic&Cloud Adoption in SME&&&\textasciitilde&&\\
\midrule
\rule{0pt}{4ex}Our Paper&Academic \& Industry&Small Business Cyber-security&\checked&\checked&\checked&\checked&\checked\\
\bottomrule
\end{tabular}\\[10pt]
\end{table}

From Table~\ref{reviewfocus}, the shortlisted reviews do not distinguish small business as a separate cohort, despite the large proportion of small businesses present in global economies. Small business is often grouped together with medium businesses in the wider cyber-security context. In this article, we examine the problems arising from such grouping. 

Alahmari and Duncan \cite{Alahmari2020} provide an overview of existing research on cyber-security risk management within the SME cohort.  They identify lack of skills/knowledge, lack of appropriate management behaviours and sub-par defence as common themes, but do not explore how such situations arise.  In this paper, we trace these symptoms back to possible contributing factors, which could enable better security solutions.

Heidt et al. \cite{Heidt2019} point out that differences between SME and their larger cousins potentially contributes to difficulties in generalising cyber-security results to SMEs. They find that data is often obtained from non-representative enterprises (e.g. larger or industry specific groups).  We re-examine the sampling issue within the industry context and find this trend extends beyond academia to industry and government.  Given the lack of industry standard terminologies, we discuss how both cyber-security self-reporting and lack of agreed definitions are contributing to confusion.

As demonstrated by Suryotrisongko \& Musashi \cite{Suryotrisongko2019}, there are a wide variety of topic areas and technologies within the study of cyber-security.  Reviews that focus on specific topics \cite{Bryan2020,Ruhwanya2019,Almeida2018,Salleh2018} provide little room for exploration of alternative solutions, e.g. a technical problem solved via a process.  In this paper, we explore cross field opportunities in the context of small business cyber-security.

In conclusion, there are no broad-based reviews exploring small business organisational characteristics that can impact their cyber-security posture, both negatively and positively.  We address these gaps in this paper.

\section{Research Data Challenges}
\label{research}
\label{sec4}

Given the importance of micro/small business in global economies, there is limited research into the cyber-security posture of, and corresponding solutions for small businesses in the 0-19 employee range.  While there is much research in large enterprise cyber-security from various countries, the transferability of this research to small business is problematic.

Each developed country is at a different stage in the small business cyber-security research journey. In the USA, CISA (Cyber-security and Infrastructure Security Agency) conducted an online survey in 2019, on providing assistance with cyber-security issues, to small businesses \cite{USSmallBusinessAdministration2019a}, but results are still awaited.  In Europe, Eurostat breaks down enterprise size across multiple variables in its most recent 2020 ICT Security in Enterprise security report \cite{Eurostat2020a}. Businesses with 0-9 employees are excluded from the above Eurostat report, highlighting the research data gap within the small business cohort. This lack of inclusion also occurs in the Australian Bureau of Statistics' business characteristics surveys \cite{AustralianBureauofStatistics8167}. These surveys ask IT and cyber-security questions, but explicitly exclude non-employers/sole traders from the respondent pool.

The cyber-security surveys conducted in Australia include:
\begin{itemize}
	\item The Australian Bureau of Statistics (ABS) annual business characteristics survey asks detailed IT usage and cyber incident questions \cite{AustralianBureauofStatistics8167}.
	\item The NSW small business commissioner CyberAware survey studied SME cyber-security \cite{NSWGovernment2017} in 2017, in collaboration with various small business related commissions around Australia. 
	\item The Australian Cyber Security Centre's (ACSC) online survey in 2019, looked at the usage of IT, security incidents and knowledge \cite{AustralianCyberSecurityCentre2020a}.
\end{itemize}

Each of the above surveys has its own focus, producing results within its own scope.  Some of these scopes overlap.  Some results appear to contradict others. For example, in answer to whether small businesses have encountered a breach/attack/cyber-crime, the percentage of people that answered in the affirmative varied based on the survey (Table~\ref{BreachResultsTable}).

\begin{table}[h]
\small\sf\centering
\caption{Small Business Self-Reported Breach Results.\label{BreachResultsTable}}
\begin{tabular}{p{3cm}p{6cm}p{2cm}}
\toprule
Source &Percentage of respondents answering affirmative/neutral for experiencing breaches&Reference\\
\midrule
ACSC Small Business Survey&62\%&\cite{AustralianCyberSecurityCentre2020a}\\
\rule{0pt}{4ex}Business Characteristics Survey&\begin{tabular}{@{}l@{}}Micro Businesses: 8.7\%\\(+17.4\% ``Don't know'')\end{tabular}& \cite{AustralianBureauofStatistics8167}\\
 &Small Businesses: 13.4\%&\\
 & (+17.1\% ``Don't know'')&\\
\bottomrule
\end{tabular}\\[10pt]
\end{table}

The discrepancies are not necessarily due to deliberate actions or errors. They appear to be a symptom of the difficulty in reaching and obtaining comparable data from the disparate and technically inexperienced cohort that is small business.  

Discrepancies in the survey results also extend past national borders. In the Hiscox report \cite{hiscox2019}, the average business cyber-security spend in European countries for organisations with 9 or less employees was US\$7,000.   In contrast, the ACSC survey \cite{AustralianCyberSecurityCentre2020a} found close to half of Australian small businesses spent less than AU\$500 annually on cyber-security.  This order of magnitude difference in expenditure from seemingly similar socio-economic cohorts requires further analysis.  While there is speculation \cite{Sommestad2014} that complexity within the cyber-security field accounts for the disparity in results, it is possible these are due to challenges in getting accurate data from small businesses.

\subsection{The Elusive Cohort}
\label{sec41}
The inability to gather data from diverse small businesses is a significant barrier to cyber-security research \cite{Valli2014,Abid2011,DeloitteAustralia2016,Gupta2005, Osborn2017}.  Consistently low response rates to broad-based voluntary surveys highlight the difficulty in obtaining a comprehensive sample.  This is exacerbated by a lack of public domain data, especially where mandatory breach reporting regulations are immature.  There is dis-incentive to self-reporting due to fear of reputational damage \cite{USGovernment2018}, given the low possibility of prosecution \cite[Fig 6.7]{UnitedNationsOfficeonDrugsandCrime2013}.

This difficulty in reaching the small business cohort could result in convenience sampling, which is known to introduce its own set of limitations and challenges \cite{Kam2007, Acar2019}.  The impact and bias arising from such sampling must be accounted for and discussed in research findings.

\subsection{Disparate Technical Terms}
\label{sec42}
Numerous industry cyber-security reports address the gap in business cyber-security \cite{Deloitte2019,PonemonIn2019, hiscox2019, Symantec2019,Vistage2018}. However, there are few standard definitions, structures and classifications within cyber-security.  Data and findings are often adhoc, incomplete and focused on headline worthy items. For example, Symantec reports \cite{Symantec2019} that globally, in 2018, small and medium businesses were more likely to be hit with ``Formjacking''. Formjacking is Symantec's term for ``use of malicious JavaScript code to steal credit card details and other information from payment forms on the checkout web pages of e-commerce sites'' \cite{Symantec2019}.  Formjacking is a subset of code injection attacks \cite{Watson2018}, where code is injected for various purposes, and not just on e-commerce sites.  The use of specialised terms such as ``formjacking'' makes comparisons between different reports difficult.

\subsection{Size Does Matter}
\label{sec43}
The differing definitions of small buisness worldwide makes data comparison difficult, see Table~\ref{TableSBDef}.

\begin{table}[h!]
\begin{threeparttable}[b]

\small\sf\centering
\caption{Small Business Definitions Around the World \label{TableSBDef}}
\begin{tabular}{p{2cm}p{2cm}p{3cm}p{3cm}p{2cm}}
\toprule
Official Business Labels&0 Employees&Micro Business&Small Business&Sources\\
\midrule
\rule{0pt}{4ex}Australia&Sole Traders&0-4&5-19&\cite{AustralianBureauofStatistics2001}\\
\rule{0pt}{4ex}UK&No-Employee Business&1-9&10-49&\cite{BritishCrown2015}\\
\rule{0pt}{4ex}European Union\tnote{1}{}&Not Applicable&0-9&10-49&\cite{Eurostat2020}\\
\rule{0pt}{4ex}US&Not Applicable&Not Applicable&$<$250 or 1,500 or industry specific annual receipt\tnote{2}&\cite{USSmallBusinessAdministration2017}\\
\bottomrule
\end{tabular}

\begin{tablenotes}
\item[1] Prior to 2020, all European Union data includes the UK.  Any data 2020 and beyond, does not include UK statistics. Where relevant, this is noted in the discussion.
\item[2] Industry dependent.
\end{tablenotes}

\end{threeparttable}
\end{table}

Each research entity adopts the local small business convention or presents a brand new definition \cite{DeloitteAustralia2016}. This disparity across the globe changes the context of findings.  Clearly, an organisation managing 200 employees operates and communicates very differently to one with less than 10 employees \cite{Cruz1997,Bales1951,Steinheider2004}.   Given the significant influence communication and human factors have on attacks such as phishing, email fraud, etc, size matters in any cyber-security discussion. 

Few researchers acknowledge the potential complexity of the large range of small business sizes.  The Ponemon Institute's data breach report \cite{PonemonIn2019} extrapolates small business data breach costs from incident costs for businesses employing 500-1000.  This cost figure is discussed as the cost for ``smaller organisations''.   The bundling of small businesses with larger (in size) cohorts hampers the ability to action the lessons learnt.
 
\subsection{Self-Reporting Fallacies}
\label{sec44}
Self-reporting of data is another problem.  The accuracy of self-reported cyber-incidents by small business owners needs examining. Non-technical persons find it difficult to report subtle cyber attacks. In addition, there are psychological self-reporting biases \cite{Donaldson2002,Fisher2000}. Thus, self-reported breaches can only be a starting point data set for analysis. 

Self-reporting surveys require business owners to be aware of cyber breaches. Common detectable symptoms of an incident, from a non-technical user's perspective, are unavailability of computer systems or data.  For example, ransomware renders a computer unusable.  Many attacks have no obvious symptoms. Active monitoring is required for subtle attacks such as data ex-filtration, man in the middle, spyware etc.  System owners may not be aware of a breach for months or even years \cite{PonemonIn2019, Redscan2019,USGovernment2018}. While large enterprises can detect subtle attacks using active traffic monitoring systems, few small businesses can afford this monitoring \cite{AustralianCyberSecurityCentre2020a}. The under-detection can be exacerbated by the lack of technical knowledge among small business owners (see section \ref{humanchallenges}).  

In Australia, the Privacy Act, the main legislation governing data, does not apply to organisations with less than AU\$3 million annual turnover \cite{OfficeofParliar2019}, i.e. most Australian small businesses. As such there are no legislative repercussions or financial incentives to report any breaches or cyber-incidents.  Embarrassment in victims of online fraud \cite{Emami2019a} is another powerful reason for Australian small businesses not reporting incidents.

Symantec \cite{Symantec2019} avoids this self-reporting gap by using known threat detection software installed within an end user's computer.  Eliminating human interaction increases accuracy by recording incident details automatically. However, the proprietary nature of this software and reporting results in only Symantec customers and devices being included. This leads to a bias towards people willing (and able) to pay for Symantec's services.  Detailed analysis and strategy design requires a more comprehensive small business sector view.

In conclusion, alternative data collection strategies are needed to supplement existing self-reporting surveys. 

\subsection{Business versus IT Perspectives}
\label{sec45}
The under-representation of non-technical respondents in online surveys deserves scrutiny. Studies targeting technically savvy businesses often reach very different conclusions to those including a broader range.

Cyber-security research relies heavily on online surveys due to logistical ease, and the assumption potential respondents use the internet for business purposes.  In contrast, less than 40\% of Australian small and micro-businesses used the internet for business development or monitoring market information in the 2017-18 financial year \cite{AustralianBureauofStatistics8167}.  

Self-selection bias is enhanced by a particular group's comfort level to a survey method. More respondents are likely to complete surveys that use familiar technology \cite{Boynton2004}. Cyber-security studies tracking industry breakdowns show IT/technical industry respondents are over-represented \cite{hiscox2019, NSWGovernment2017, Osborn2017}.  

Challenges exist around demographic representation in samples with online surveys as the sole method of data collection. ``Younger, male, avid Internet users, and those with greater technological sophistication'' \cite{Kwak2002} are more likely to complete online surveys compared to surveys done via traditional mail. In contrast, in 2016, over half of small business owners in Australia were over 45 \cite{CommonwealthofAustralia2019}.  

A survey with more technically literate respondents can introduce significant biases in results. Studies with input from technical experts \cite{Dojkovski2010, Coertze2013, Panjwani2017} identified different concerns compared to general small business owners \cite{Bada2019a, Dimopoulos2004, Almubayedh2018, Laleh2013}.  A series of interviews from a single study \cite{Osborn2018} shows that small scale IT users have constraints in security decision making that are not expected by technical security providers. Australian small businesses in information, media and telecommunication, and professional, scientific and technical services categories made up less than 14\% of all businesses, by business count, in 2019 \cite{AustralianBureauofStatistics2020b}.  

In summary, the majority of small business owners in Australia cannot be assumed to intuitively understand the technical fundamentals central to cyber-security.  Exclusively sourcing data from voluntary online surveys favours respondents comfortable with technology.  Results from surveys must account for and reconcile any gaps between sample respondents versus small business profiles, from both demographic and technology literacy perspectives.

\section{Challenges Faced by Small Businesses}
\label{constraints}
\label{sec5}
Large enterprises have been amongst the earliest adopters of cyber-security technologies.  Current business cyber-security solutions favour these businesses in terms of scale, cost and usage. 

Today's cyber-security lessons and conventions are a result of early large-scale incidents e.g. NotPetya, Equifax, Wikileaks etc affecting mostly large organisations. With new threats constantly emerging, large enterprises are ideal customers for vendors, with a higher likelihood of return on investment. Consequently cyber-security industry best practices, standards and products are heavily influenced by the needs of larger organisations.  

Small business cyber-security cannot be ``cut and paste'' from large scale solutions.  Resource availability, technical landscape and operational processes of small business need careful consideration. Small businesses face some common challenges that need to be accounted for in cyber-security strategies since cyber-criminals are looking for smaller targets \cite{Symantec2019,Smith2016,McDonald2019}.

\subsection{Technical Challenges}
\label{techchallenge}
\label{sec51}
The technical landscape of a small business can potentially be very different to that of a large enterprise \cite{Osborn2017}, making it impractical to apply solutions for the larger enterprises to smaller scale users. The small business IT technical landscape (``architecture'') in which the cyber-security solution must function is a barrier to adopting a solution.  

Current industry practice tests changes in an environment separate from customer-facing sites and systems to ensure the changes do not impact IT system availability for live customers.   A test or staging environment is a copy that mirrors the customer-facing (“live”) business system in hardware, software and configuration. Testing in a separate environment is recommended by many application providers before software upgrade processes \cite{Microsoft2017a, Raul2020}.   This testing ensures unintended consequences of changes/tests do not affect the business' live systems, keeping them ``safe'' and operational.  A test environment can be used to conduct destructive testing such as disaster recovery scenarios.

Cyber-security solutions designed to test a response to debilitating events require a safe testing environment. For example, denial of service (DoS) simulation tools \cite{Iyamuremye2018, Rapid72019} simulate a service overwhelmed with requests resulting in legitimate requests not getting through.  A DoS simulator, if implemented on a live system, would render the business IT infrastructure, e.g. website, unavailable to customers, or worse, jeopardise overall system integrity. Live environments cannot be used for stress-inducing tests. Consequently, businesses without a test environment will never be able to test the full suite of catastrophic scenarios as part of their incident response training.

A test environment requires substantial technical knowledge, time and ongoing maintenance.  Australian and New Zealand small and medium enterprises spend only 6\% of their total revenue \cite{DeloitteAustralia2016} on IT.  For 2016, the median Australian small business had an average turnover of AU\$125,000 \cite{CommonwealthofAustralia2019}, translating to an annual IT spend of about AU\$7,500.  This budget covers all IT spend, including hardware, software, network services, IT personnel and other incidentals, making a test environment just one of many competing priorities.  While cloud technologies have reduced costs associated with test environments, the technical knowledge required is still substantial.  Less than a quarter of Australian small businesses have in-house, qualified IT support personnel (with roughly a third using contractors/consultants as needed) \cite{AustralianBureauofStatistics8167}.  There is no mention of small business testing environments for off-line testing in the literature.

The small business move towards cloud infrastructure and services makes large enterprise cyber-security solutions unsuitable. Over a third of micro businesses and half of small businesses used cloud based services in 2018 \cite{AustralianBureauofStatistics8167}. This demand is expected to increase \cite{Deloitte2019a} and accelerated due to the COVID19 pandemic \cite{GartnerInc2020}.  While official statistics on cloud services adoption during the pandemic are still pending, anecdotally the uptake of third party IT services in general has increased substantially \cite{Chiappetta2020}, many of which are enabled through the cloud \cite{Dignan2019}.  This move away from private, or on premise, infrastructure (i.e. IT equipment set up and owned by the business) is driven by the associated set-up and maintenance costs \cite{Deloitte2019a}, a proposition especially attractive for resource constrained smaller businesses. 

Cloud infrastructure presents challenges for traditional cyber-security products. For example, general network scanning \cite{Detken2013} assumes the infrastructure is local and/or reachable for scanning purposes, an assumption valid only for private infrastructure.  The increasing suite of cloud-based network perimeter scanning services  \cite{Spyse2020,Tenable2020,Intruder2020} require explicit input from users on the target devices to be scanned. Targeted device scanning is a subset of a complete network scanning solution and cannot discover private assets (e.g. printers) with external interfaces left accidentally unsecured.  This potential gap could be addressed by other technical or process remedies, and needs further examination by individual businesses - a task requiring technical knowledge and resources that many small businesses do not have.

Scanning in the cloud is a challenge for shared computing resources.  Any customer sharing the same target computing resource will be affected if the targeted application is overwhelmed. Cloud infrastructure providers impose strict conditions on security-related testing activities \cite{AmazonAWS2020,Microsoft2020}, with the notable exception being Google Cloud \cite{Google2020}.  


Nearly 90\% \cite{AustralianBureauofStatistics8167} of small business cloud users utilise software in the cloud.  Many application providers explicitly prohibit activities interfering with their service, preventing customers from pro-actively discovering any vulnerability in their service provider. Indemnity and liability around use of an outsourced function needs clarification \cite{August2011, Selznick2018}.

It is necessary to address when testing and monitoring of privately owned devices is permissible.  Mixed-use personal and company-owned assets, e.g. mobile devices, are common amongst small businesses \cite{ACMA2013,Almubayedh2018, Osborn2017}. Mobile device management (MDM) software \cite{Harris2014,Google2020a} allows businesses to monitor data and software residing on and transiting through devices. MDM can freeze certain functions and/or data, and, if needed, reset the entire device.  Many large enterprises routinely deploy MDM on devices issued to staff.  The level of access that MDM requires raises questions of the extent of non-work information shared with work IT administrators \cite{Perez2019} when a device is not a work exclusive device.  

To address the privacy concerns of MDM, mobile application management (MAM) offers a subset of the functions of MDM \cite{Microsoft2017,Google2020b}.  MAM functions require business trade-offs and must be paired with mitigating policies based on the sensitivity of business data. This requires expert guidance and is at odds with other revenue generating priorities for a small business.

Clearly, re-using cyber-security solutions designed for large scale infrastructures is not a feasible strategy for small businesses. More appropriate cyber-security products are needed for small business given their mix of stand alone and shared IT resources. The liability arising from small business' use of cloud service needs clarification. 

\subsection{Human Challenges}
\label{humanchallenges}
\label{sec52}
Small business cyber-security human resources are very different to those of large enterprises.   A qualified and accountable IT department is common in large enterprises but not in small businesses.  

In a European \& US survey \cite{Dimopoulos2004} less than 30\% of small businesses, and less than 5\% of micro businesses, indicated existence of a security administrator or any formal IT security qualifications.  Approximately 10-25\% of Australian micro and small businesses have no IT support at all \cite{AustralianBureauofStatistics8167}.  

In Australia, most small business support comes from external contractor/consultants, suppliers of software/hardware or non-IT qualified staff \cite{AustralianBureauofStatistics8167}. External contractors are engaged for specific tasks. They do not undertake cyber-security risk assessments unless contracted specifically. Suppliers only support supplied software and/or hardware, unless an extra support contract has been purchased.  Small businesses' constrained resources and the low number of devices used imply support contracts are unlikely.
  
Cyber-security requires risk analysis and mitigation of the entire IT system, not just individual components. The small IT budgets of small businesses do not cover the salary of an IT administrator \cite{Glassdoor2019}. 

Furthermore, small business owners are at a knowledge disadvantage in advocating for their cyber-security needs. In the start-up phase of a small business, owner-managers have to perform any functions that they cannot afford financially, or haven't had the time, to outsource or hire. As discussed previously, given small business' small IT budget, these include being the IT support person, and by extension the cyber-security person.  Some tasks, such as cleaning, physical security etc. can conceivably be done by a person with limited experience. A cyber-security assessment cannot be done effectively by a novice. 

The age demographic of Australian small business owners indicates less exposure to technology.  Nearly 60\% of Australian small business owners were born before 1971 \cite{CommonwealthofAustralia2019} - well before publicly available Internet \cite{Taylor2014}.  This technical skills gap is reflected in OECD testing \cite{OECD2019} with less than 40\% of Australian adults scoring within levels 2 or 3 (out of maximum 3 levels) of OECD's computer problem-solving skills scale. This percentage falls to 17.2\% within the 55-65 age group.  

The timing of technology introduction places Australian business owners at a disadvantage when dealing with cyber-security.  Assistance must be designed taking into account this lack of experience and cater to non-technical small business owners.  However, training in a receptive format is well received by small businesses \cite{Walker2007}.  In section~\ref{agility} we discuss the advantages inherent in this demographic.

\subsection{Organisational/Process Maturity}
\label{sec53}
Cyber-security requires ongoing effort in education, process and investment and is effected by the IT maturity of organisations. In Australia, approximately a third of small businesses operating in June 2015 did not survive to June 2019 \cite{AustralianBureauofStatistics2020b}.  The small business sector constantly has new entrants.  Small businesses in early inception and survival stages do not focus on processes \cite{Scott1987}. In these early phases, owners/managers usually have direct oversight of business tasks with formal processes coming in as the business matures and expands.  This is at odds with a well-rounded cyber defence posture \cite{Alqatawna2014} requiring business policies and controls.  This lack of clarity on process details (for example on IT fix times and specific responsibilities around patching) leaves the organisation exposed to opportunistic cyber-criminals.  

Research on small business behaviour in the early stages of business is required. Any cyber-security solutions or policies for small business must consider that the business' focus may not be on the management of cyber-security processes and rules.

\subsection{Industry Standards}
\label{sec54}
The bewildering variety and complexity of industry standards is a challenge that small business owners face. Some examples of cyber-security standards and frameworks include:
\begin{itemize}
	\item NIST \cite{NationalInstituteof2018}
	\item COBIT \cite{isaca2019}
	\item ISO27001 \cite{ISO27001}
	\item Australian Government Information Security Manual \cite{AustralianSignalsDirectorate2020a}
\end{itemize}
When used by a technical stakeholder such as a Chief Information Officer, these guidelines and standards fit any business wishing to protect itself from cyber attackers.  The base assumption of many standards is a technically literate person being available to guide the business through analysis and implementation.  Few small businesses can justify such a resource especially in the initial phase of setting up the business.  Simplified versions of the above standards are available \cite{CommonwealthofAustralia2017}, for example, the Australian Signals Directorate's (ASD) Essential 8 \cite{AustralianCyberSecurityCenter2019} and the small business cyber-security guide \cite{AustralianCyberSecurityCentre2020b}.  However, the range of available summarised resources vary in depth and format.  More importantly, these resources have limited comparability between each other or existing security standards.  It is unclear how a small business' cyber-security work fits in with other advice and standards once one list is completed. 

The complex relationships between the different standards leaves non-technical small business owners with a very low sense of self-efficacy, or sense of control.  The relationship between self-efficacy and action is a well-established driver of human behaviour \cite{Maddux1983}. In addition, low self-efficacy in cyber-security could be a factor in an owner rationalising inaction \cite{Renaud2016}.  

Industry bodies have recognised the difficulties faced by small businesses.  ISACA, the governing body for COBIT, indicates small and medium enterprises will be catered for as part of their COBIT 2019 standards  \cite{ISACA2019b}.  Unfortunately, to date, this focus area is still under review and not available. 

Until cyber-security standards recognise small business' constraints, small business will find it hard to contextualise cyber-security within their own business, or implement the necessary controls.

\subsection{Cyber Insurance} 
\label{sec55}
Cyber insurance is a challenge for smaller businesses.   Traditionally, insurance policies are used to protect a business against disruption or loss, ranging from natural disasters to theft. Business cyber insurance is a relatively new phenomenon, with many variations offered by different providers.  

Various providers in Australia offer cyber insurance to their small business customers \cite{InsuranceBusinessAustralia2020}.  Each insurer appears to require slightly different levels of due diligence from the small business, while each policy provides coverage based on disparate internal standards and policies \cite{CyberInsuranceAustralia,BizCover2019,Davenport2017}.  Further study and research is needed to provide clarity to small businesses around standards applied across insurers.

Cyber insurance coverage has been called into question.  One highly publicised debate \cite{Satariano2019,OECD2017} surrounds the insurance exemption for ``act of war'' or ``terrorism''. In most politically stable countries, a clause to exclude coverage of catastrophic events, e.g. war/terrorism is employed to manage underwriting exposure \cite{OECD2017}.  Most insurance policy takers usually accept an act of war in a developed country to be a remote, acceptable business risk.  

The effectiveness of such cyber insurance came under scrutiny when Mondelez submitted a claim on suffering US\$700 million in damage from NotPetya malware in 2017 \cite{Satariano2019}.  Zurich Insurance denied the claim, citing the attack as an act of terrorism, as NotPetya was designed to cause maximum damage as opposed to financial gain.  The situation was further exacerbated when the US government attributed NotPetya to Russian origins. The ongoing conflict between Russia and Ukraine further strengthened the terrorism case as Mondelez had infrastructure in Ukraine.  The decision by Zurich continues to be challenged in court by Mondelez, and other insured companies.  The issue of terrorism is so contentious that Lloyd's of London \cite{LloydsofLondon2017} explicitly excludes terrorism related scenarios in its cyber insurance cost analysis.

``It is difficult to attribute a cyber-attack to a particular group or actor'' \cite[P.17]{LloydsofLondon2017}.  The difficulty stems partly from code from one malware author being reused by another.  When malware code is examined forensically, it may contain artefacts of authors other than the criminal.  For example, the NotPetya malware contains code stolen from the US National Security Agency (NSA) \cite{Greenberg2018}. Thus the malware contains signatures indicative of US origin, despite the attack being attributed to Russia.  Accurate attack attribution is out of reach of the average organisation, especially small business.  This lack of attribution makes it difficult to prove an insurance claim as non-terrorism.

Small businesses and personal computers are often collateral damage of large scale cyber-attacks.  The NotPetya malware wiped clean 10\% of all computers in Ukraine \cite{Satariano2019}, along with many other international companies (including bouncing back into a company in Russia itself \cite{Greenberg2018}).  Many small businesses were highly likely to have been caught up in this fallout. Numerous Australian SME and large corporations were impacted by the 2016 Petya attack \cite{NSWGovernment2017}, from which the NotPetya malware is derived.

In addition to the Mondolez owned Cadbury factory in Hobart, NotPetya affected companies with an Australian presence such as DLA Piper, Maersk, TNT etc \cite{ABCNews2017, Crozier2018}. Unfortunately, following the Australian government's attribution of NotPetya to Russia, and NotPetya's classsification as malware \cite{Taylor2018}, any Australian businesses' with an act of terrorism insurance exception would reasonably find themselves not covered by their cyber insurance policy.

The battles still being fought through the international judicial system indicate that cyber insurance is not mature enough for Australian small businesses to use effectively. For cyber insurance to be a risk mitigation tool for small business, significant industry, regulatory and assistance efforts are needed to clarify and provide adequate coverage to small businesses in case of a breach.

\subsection{Legal Remediation}
\label{sec56}
Despite the act of breaking into computers being covered under various legislations in different countries \cite{UnitedNationsOfficeonDrugsandCrime2013}, cyber-criminals continue to operate with impunity. One reason for this perceived lack of consequences is the legal complexity of investigating and prosecuting cyber-crimes. 

Criminal and remediation processes are complicated when the pepetrators are under a different legal jurisdiction to the victims and properties and hence, subject to different legal processes and laws.  The ease with which attacks can be launched over the internet makes cross-jurisdictional cyber-crime possible for criminals with minimal resources.  Over half the countries responding to a 2013 United Nation's (UN) survey described a transnational element in the majority of the cyber-crimes reported \cite[P.117]{UnitedNationsOfficeonDrugsandCrime2013}. 

This cross jurisdictional complication is discussed in the 2014 Australian government \cite{Cross2014} study on online fraud. The negative impact of transnational barriers on investigations and judicial outcomes for victims continues today \cite{Bainbridge2020a}. Jurisdictional complications possibly contribute to the overall low conviction rate (10\%) of police recorded cyber-crimes \cite[P.172]{UnitedNationsOfficeonDrugsandCrime2013}. However, international co-operation in fighting cyber-crime is beneficial \cite{Reitano2015} in overcoming structural barriers to prosecuting international cyber-crime.

The low conviction rate is exacerbated by the low rate of reporting.  Most cyber-crimes are not likely to be reported \cite[P.119]{UnitedNationsOfficeonDrugsandCrime2013}.  In Australia, the top 3 reasons \cite{Emami2019a} for non-reporting of online fraud include embarrassment, belief police cannot find offender and uncertainty of right reporting authority. Given these factors, the overall redress that actually occurs is even smaller.

Even in the remote event a cyber-crime results in a conviction, the loss of confidentiality of data is a major issue.  Once a receiver has stolen data, this knowledge/data cannot be removed from the receiver.  Even where data can be located, the ease with which copies can be made and distributed makes containment difficult. One such example is the classified data taken from the United States National Security Agency (NSA) by an insider \cite{Bamford2014}. Despite investigations and knowledge of the locations of the document caches, information from this breach is still available on the Internet \cite{Szoldra2016}.

Outside of legislative penalties, there is little incentive for Australian small businesses to report cyber-crime.  With low remediation rates and investigative difficulties, reporting to authorities is replaced by the priority of recovering from the cyber-incident.  Without dedicated legal personnel to follow up and navigate the complex systems involved, legal remediation remains out of reach of most small businesses. 

Substantial work is still needed on the cyber-crime legal framework to convince small businesses that it is effective in delivering remediation when needed.

\subsection{Cost of a Data Breach}
\label{sec57}
The gap in knowledge of the cost of inaction makes investing in cyber-security a difficult decision for small businesses.  A small business needs to weigh up each investment decision against perceived benefits/loss. Existing data breach costing studies around the globe have primarily focused on larger businesses \cite{Deloitte2019,USGovernment2018,PonemonIn2019}.   

Small businesses (1-49 employees) suffer an average loss of US\$14,000 per year per firm \cite{hiscox2019} across the US and some European countries. Despite a low small business sample size, a US Government report \cite{USGovernment2018} speculates the impact/cost of a breach on small business is possibly more devastating than for a large business due to the loss of customers. There is no data available for Australian businesses in the 0-19 employee range.  In a survey by ACSC \cite{AustralianCyberSecurityCentre2020a}, about half of small business owners predicted it would take a few days to recover from a hypothetical cyber attack.  

Until relevant cyber-breach cost data is available, it will be hard to persuade small business owners to invest in managing cyber-security risks \cite{AustralianGovernment2016}.  Scientific and industry research is needed to examine ways of determining actual costs to small businesses from cyber breaches, similar to larger enterprises \cite{USGovernment2018,PonemonIn2019}.


\section{Advantages For Small Businesses}
\label{opportunities}
\label{sec6}

The small business cohort has advantages over larger enterprises, despite the many challenges they face.  With the right assistance, a small business can become a less attractive target to cyber-criminals.

\subsection{Small Business Agility}
\label{agility}
\label{sec61}
Two advantages for small businesses over their larger counterparts are their flexibility and willingness to learn.  The COVID-19 pandemic showcased the agility with which small businesses adapted to fluid situations \cite{Khadem2020,Broadsheet2020,Gallagher2020,Longbottom2020,Gray2020}.  
Small business owners possess qualities allowing them to react with agility \cite{DeVries2006,Branicki2018}.

The older demographic of small business owners comes with advantages. Creative and critical thinking related digital skills actually improve faster within older age groups with this improvement attributed to everyday hands-on experience of working with technology \cite{EshetAlkalai2010}.

Flexibility, willingness to learn and creative thinking are  important cyber-security skills \cite{Freed2014,Bashir2015,Hall2020}. However small business owners, dominated by older but adventurous learners \cite{Zhao2006,Moran1998}, still struggle with cyber-security \cite{Laleh2013,Renaud2016}. Creative thinking skills cannot be used effectively when the basic information needed to create solutions is unavailable in an accessible format. Unlike conventional IT knowledge, e.g. productivity software and mass-market hardware, cyber-security knowledge is still highly technical and inaccessible to the public.  In Australia, small business owners are largely offered adhoc cyber-security workshops and resources via disparate small business training and communication channels, e.g. \cite{CommonwealthofAustralia2017,BusinessVictoria2019, AustralianCyberSecurityCentre2019e}. With research showing that cyber-security needs to be a gradual, ongoing and persuasive effort \cite{Vroom2004,VonSolms2013}, alternative formats and continuous approaches to small business cyber-security training need to be examined.  

A targeted capacity building curriculum, encompassing both prevention and response, can be used to take advantage of small businesses' agility.  Building practical coping skills appeals to creative problem solvers like small business owners, without relying on formal processes \cite{Branicki2018}.  A good starting point would be cyber-security advice for small businesses focusing on prevention, e.g. password policies, spotting phishing etc.  This advice should extend to other aspects of a cyber-incident's lifecycle such as incident response, record keeping, technical recovery strategies etc.  

Victim assistance facilities (as part of incident response) like IDCare \cite{IDCare2019}, ACSC Report \cite{AustralianCyberSecurityCentre}, Scamwatch \cite{ACCC2019} etc do exist in Australia but public awareness is not ubiquitous \cite{Emami2019a}. Increased education around crime reporting processes and forensic practices will improve response to breaches by small business owners.

Further research is needed to enable cyber-security education to take advantage of small business owners' agility.  Solutions need to be tailored to \cite{Walker2007}, and take advantage of, the unique abilities of small business owners.

\subsection{Small Business Networking and Alliances}
\label{sec62}
An alliance between small business owners would enable data collection and advanced intelligence sharing, as well as peer support. This real time knowledge of trending threats and attack methods would help small businesses to be vigilant to emerging threats.

International large enterprises \cite{Verizon2019,Abolhassan2017} have formed cyber-security alliances allowing members to gain advanced intelligence and work collaboratively, and forming a cyber-security feedback loop. In Australia, alliances between industry, government and research such as AusCERT \cite{AusCERT2019} and ACSC Partners \cite{AustralianCyberSecurityCentre2021} exist. Unfortunately these alliances target organisations with dedicated cyber-security IT personnel/budgets. For small businesses such an alliance would provide easily accessible intelligence and affordable skills. \cite{Yanpiboon2017}.

A cyber-security alliance of small business owners could allow law enforcement, legislators and technical support companies to respond rapidly to emerging threats.  Each small business could contribute cyber-security incidents and intelligence into a central database providing cyber-security peer support to each other.  Education and awareness act as extensions of such an alliance, leading to it becoming a trusted authority on small business cyber-security.  

Technical and administrative facilitators are needed to overcome existing communication barriers between small businesses and facilitate such an alliance.  The exact facilitation and support needed to start and maintain such a new alliance needs to be researched and clarified.

\subsection{Advantages of Zero Trust Model}
\label{sec63}
Suitable security models differ due to the scale of organisations. The low requirement for IT homogeneity makes zero trust security models \cite{Rose2020} a prime framework for small business.  Small businesses have limited large scale technical legacy to undo and their scattered IT operations suit a zero trust model. 

The traditional network perimeter safeguards an IT eco-system that is relatively obstacle-free internally, leading to the analogy of a hard shell enclosing a chewy centre \cite{Kindervag2016}. The problem of an attacker having free reign once a perimeter is bypassed has been demonstrated in large scale breaches \cite{115thCongress2018,AustralianNationalUniversity2019}. 

Google \cite{BeyondCorp2014}, as well as other stakeholders \cite{Curran2016}, have recognised the futility of safeguarding an ever changing network perimeter \cite{Ward2014,SERadio2019}, and have adopted zero trust principles in their services in their BeyondCorp design.  The zero trust security model \cite{Kindervag2010} is a network model for defence that presumes an attacker has gained access inside the hardened perimeter.  In a zero trust model, access control is no longer performed at the network surrounding the IT systems, but focused on authentication and authorisation of user access to individual services.

Small businesses have difficulty with perimeter defence as they don't own many of the IT services used. In many ways, the zero trust architecture is one small businesses have been managing for many years.  Zero trust allows focus on individual service protection, thus limiting impacts even if an individual service gets breached.

However, small businesses need assistance with implementing a zero trust model. Each new device needs arduous processes of set-up for multiple services rather than one perimeter \cite{Escobedo2017}.  Abstracting security into a layer/functional component \cite{Cittadini2016} allows standardisation and re-use across multiple businesses. This creates an opportunity for a reusable design/platform that small businesses can implement in their unique IT environments.  

Currently there are very few products and little research in this context. Despite implementation within large international companies such as Google and Akamai, zero trust has yet to see significant momentum within the Australian business landscape. Development of zero trust IT ecosystems for small business could bring high returns due the large number of end users.

\subsection{Open Source Cyber-security Tools}
\label{sec64}
Open source software \& hardware (OSSH) is developed and (usually) released for free by a community of software makers and hardware designers.  OSSH is an option used by IT professionals when paid IT products, including cyber-security products, do not meet technical or cost requirements. In embracing OSSH, Australian small businesses would be following in the footsteps of Australian government authorities.  Various Australian government departments have encouraged use of OSS software as well as making software open source in general \cite{QueenslandGovernment2020,DigitalTransformationAgency,VictorianGovernment,AGIMO2011}. However, a lack of technical expertise has hampered small business' ability to take advantage of free OSSH security solutions \cite{Cisco2020,Welte2014,Lyon2019}.   

The ethos of open contribution by volunteers results in OSSH products being released under broad licenses (e.g. GNU general public use \cite{FreeSoftwareFoundation2007}, creative commons license\cite{CreativeCommons2020} etc.) that permit most uses. In some cases, this includes commercial use with nothing more than attribution.  OSSH has produced popular projects such as the MySQL database, Fedora Linux operating system, Arduino microcontroller etc which enable the development of commercial products.  A variety of free OSSH cyber-security tools are now available \cite{CyberX2018}.

However, characterising OSSH as free or low cost ignores the overall cost of ownership beyond initial financial outlay for supporting equipment and setup effort \cite{Ven2008}.  Both the Equifax breach \cite{115thCongress2018} and industry standards like ITIL \cite{Alojail2013} have demonstrated the critical role played by post-implementation operational support in securing an IT environment.  

To enable ongoing maintenance, enterprise versions of OSSH are sold by third party vendors including items such as on-going support, updates, certifications etc. For example,  RedHat Enterprise Linux is based on OSSH Fedora Linux \cite{RedHat2020} and offers the support options that Fedora does not. In larger businesses, where enterprise level OSSH products may not exist, in-house support is enabled by internal staff with dedicated responsibilities.  A lack of technical expertise and financial resources in a small business is an impediment to the use of OSSH.  

OSSH presents an opportunity for technical security solutions for small business owners.  OSSH provides many benefits for small businesses including:
\begin{itemize}
	\item On-going reviews minimising security vulnerabilities from human error \cite{Ebart2008}.
	\item Reducing vendor lock-in \cite{Ven2008}.
	\item Being part of the software community with a stake as users in the software development \cite{Wesselius2008}.
	\item Encouraging good cyber-security hygiene habits by enabling adoption in the early stages of the business lifecycle.
\end{itemize}
Australian small businesses could leverage the expertise of developers active in open source development. They could participate in local open source communities like PyCon \cite{PyConAU2021} and Linux Australia \cite{LinuxAustralia} to make their business needs met and their voices heard. Building bridges between the OSSH and small business communities could provide value for both.

Given the importance of cyber-security in business, further research is needed to analyse the suitability of OSSH security products into the Australian small business landscape and the obstacles to be negotiated to remove entry barriers. 


\section{Discussion}
There are few reviews dedicated to the small business cyber-security landscape. Our research reveals many constraints in small business adoption of cyber-security practices, but also advantages in the cohort that should be leveraged.

\subsection{Data Collection}
As Table~\ref{datacollectionopp} illustrates, data collection in small business cyber-security research requires improvements in regards to scope, consistency and respondent range. 


\noindent\begin{minipage}{\linewidth}
\captionof{table}{Cyber-security Data Collection Practice Issues and Desired Outcomes}
\label{datacollectionopp}
\begin{tabular}{p{1cm}p{5cm}p{5cm}p{2cm}}
\toprule
Section&Current Challenge&Proposed Solution&Proposed Responsible Bodies\\
\midrule
\rule{0pt}{1ex}~\ref{sec4} \& ~\ref{sec41}&Lack of focus \cite{Eurostat2020a,AustralianBureauofStatistics8167} on micro businesses in research.&Focused research on $<$20 employee businesses to capture the operational behaviour of small businesses.&Government; Research\\
\hline
\rule{0pt}{1ex}~\ref{sec42}&Use of novel technical \cite{Symantec2019} terminologies.&Standard terminologies used to ensure language describes the same phenomenon.&Standards Bodies; Industry\\
\hline
\rule{0pt}{1ex}~\ref{sec43}&Use of disparate \cite{DeloitteAustralia2016} grouping sizes.& Acknowledgement of group size range impacts in findings.&Industry; Research\\
\hline
\rule{0pt}{1ex}~\ref{sec44}&Unaccounted self-reporting bias \cite{Donaldson2002,Fisher2000} in results.&Social desirability and awareness adjustments incorporated into self-reported statistics to account for bias from collection method, both technical and psychological.&Statistical Collection; Research\\
\rule{0pt}{3ex}&Over reliance on surveys \cite{AustralianCyberSecurityCentre2020a,NSWGovernment2017,AustralianBureauofStatistics8167} as a data gathering instrument.&Detection based small business monitoring to reduce reliance on self-reported statistics.&Industry; Government\\
\hline
\rule{0pt}{1ex}~\ref{sec45}&Survey instruments \cite{Boynton2004} favouring technically savvy  \cite{hiscox2019, NSWGovernment2017} cohorts.&Demographically representative respondents of the small business landscape sourced to obtain relevant understanding of gaps and priorities.&Government; Statistical Collection\\
\hline
\multicolumn{4}{p{\textwidth}}{{\footnotesize Note: In the rapidly evolving field of cyber-security, research is advanced by a combination of different parties: scientific, government, industry, journalistic etc.  The``research'' bodies referenced here, and in subsequent tables, include all who investigate the cyber-security challenge in question.}}\\
\bottomrule
\end{tabular}
\end{minipage} 

From our review of the literature, current cyber-security research lacks representative results and insights on small business only cohorts due to being bundled with better resourced medium businesses.  Comparison between widely disparate groups introduces difficulties in producing focused learnings that can be applied to micro businesses and sole traders.  Further research is needed on the sub 20 employee cohort to produce targeted findings. 

Our research shows that the wide variety of language and definitions used within the cyber-security industry poses issues with interpretation of survey/research findings. Mismatched cohort sizes make comparisons difficult, and sometimes misleading.  Standardisation of both group definitions and terminologies is needed within cyber-security research and industry to enable useful comparison and longitudinal use of research data. 

The heavy use of self-reported surveys introduces social desirability and awareness bias in results.  Rather than relying solely on self-reported data for small business research on quantitative matters e.g. breach rate, technology involved, we suggest a wider deployment of detection based collection.  For qualitative research, researchers should actively account and compensate for self-reporting bias, if possible, as part of their results.

Finally as our research highlights, distortion in research results is worsened by the technical methods by which many surveys are conducted.  Technical data collection channels such as online surveys, result in samples biased towards respondents comfortable with using technology. Demographic data shows small business owners are not all technically savvy.  Surveys need to be accessible and promoted towards both technically savvy and non-technical respondents to ensure all types of business owners are represented.

\subsection{Small Business Challenges}
In addition to the research challenges listed above that being a small business poses many hurdles in understanding and implementing cyber-security, as summarised in Table~\ref{smallbusinesschallenge}. 
\pagebreak


\label{smallbusinesschallenge}
\captionof{table}{Cyber-security Challenges for Small Businesses}
\label{smallbusinesschallenge}
\begin{longtable}{p{1cm}p{5cm}p{5cm}p{2cm}}
\toprule
Section&Current Situation&The Way Forward&Best Responsible Bodies\endhead
\midrule
\vspace{1cm}
\rule{0pt}{1ex}~\ref{sec51}&Lack of clarity on application user liability \cite{August2011,Selznick2018} in the event of a breach.& Clarification of breach liability if small business uses an external IT service.&Professional Bodies; Government\\
\vspace{1.5cm}
\rule{0pt}{1ex}&Small number of cyber-security solutions that take into account mixed physical ownership \cite{ACMA2013,Almubayedh2018} IT environments.&Small business mode of operation research and product development that recognises the need for hybrid IT environments.&Research; Industry\\
\vspace{1.5cm}
\rule{0pt}{1ex}&Effective reuse of current cyber-security solutions are prevented by existing differences between large and small enterprise IT architectures [Section~\ref{techchallenge}]. &Small business common IT architecture research and tailored resources built for small businesses leading to fit-for-purpose cyber-security tools and advice.&Research; Government\\
\hline
\rule{0pt}{1ex}~\ref{sec52}&Small business owners have several challenges in technical knowledge to effectively use cyber-security information today.&Development of material suitable for small business knowledge level and mode of learning&Government\\
\hline
\rule{0pt}{1ex}~\ref{sec53}&Small businesses are likely to be in an inception state \cite{AustralianBureauofStatistics2020b} with immature processes and organisation \cite{Scott1987}.&Further development of standards and tools that takes organisational immaturity into account.&Industry; Research\\
\hline
\rule{0pt}{1ex}~\ref{sec54}&Industry standards \cite{NationalInstituteof2018,isaca2019,ISO27001,AustralianSignalsDirectorate2020a} require a high level of technical expertise to follow.&Standards that are targeted towards a small business audience recognising external constraints \cite{ISACA2019b} with a migration path to existing full standards.&Standards Bodies; Research; Government\\
\hline
\rule{0pt}{1ex}~\ref{sec55}&Lack of clarity \cite{Satariano2019} around insurance requirements and coverage.&Further research, product development and guidance for small business on all aspects of cyber insurance to enable informed choices and adequate coverage. &Research; Government\\
\hline
\rule{0pt}{1ex}~\ref{sec56}&Lack of small business self-efficacy around legal remediation \cite{UnitedNationsOfficeonDrugsandCrime2013} from cyber-crime.&Further clarification, research and development around legal aspects of cyber-crime and associated remediation for small businesses.&Government\\
\hline
\rule{0pt}{1ex}~\ref{sec57}&Lack of relevant data \cite{USGovernment2018} on costs associated with cyber breaches for small business.&Further research to clarify the cost of data breaches for comparable small business cohort, to enable accurate risk based decisions making from small business.&Research; Statistical Collection\\
\bottomrule
\end{longtable}

Our research found that small businesses tend to operate differently from large corporations due to their size, leading to different IT usage pattern. One phenomenon is the tendency to mix personal and business use in devices and securing these devices using traditional security solutions such as MDM and MAM pose ethical and logistical dilemmas. The rising use of cloud services by small business also raises questions around liability and the control and visibility a small business actually has over its IT security.  

Through demographic statistics we found that small businesses and their owners are at a disadvantage in regards to experience with and knowledge of technology.  Given the older demographic, the majority of small business owners have spent a smaller proportion of their working lives on the internet, when compared to younger generations. This lower level of technical literacy needs to be accounted for in training and standards.  It is unrealistic to expect non-technical people to self-drive and implement the highly technical security standards available today.   

Small businesses' tendency to be at an inception stage is also at odds with the process and oversight driven cyber-security standards, with research showing early stage businesses do not have many established processes and management practices.  Standards and solutions for small businesses need to factor in this mode of operation.

The general lack of maturity around cyber insurance, judicial processes and cost data all contribute to the state of confusion in cyber-security for small businesses.  The ongoing litigation between insurance providers and larger corporations shows cyber insurance is not yet an effective risk management tool for small business. Furthermore, the lack of comparable cyber-incident data leaves business owners with little understanding of the potential loss they are insuring against.  Compounding this are the general difficulties in successfully prosecuting cyber-criminals, leaving small business owners with a lack of self-efficacy in the judicial system.  The low sense of self-efficacy leads to inaction and is reflected in the low rate of reporting.  Unfortunately, a low rate of reporting leads to cyber-criminals continuing to operate with impunity.

Our examination reveals the need to better understand small business IT usage, demographics, mode of operation and structural constraints.  By understanding the differences and challenges involved, the discussion around cyber-security could match small business users' needs and expectations.

\subsection{Opportunities}
By taking a step back to look at small business as a whole, our research (summarised in Table~\ref{smallbusad}) found opportunities for small business with regards to cyber-security. These opportunities take advantage of the unique landscape and characteristics of small businesses and their owners, and if leveraged, would lead to more resilient small businesses.

\noindent\begin{minipage}{\linewidth}
\captionof{table}{Small Business Cyber-security Advantages}
\label{smallbusad}
\begin{tabular}{p{1cm}p{5cm}p{5cm}p{3cm}}
\toprule
Section&Advantages&Opportunities&Best Responsible Bodies\\
\midrule
\rule{0pt}{4ex}~\ref{sec61}&Agility and resilience of small businesses \cite{DeVries2006,Branicki2018} in the face of challenges - take advantage of traditional flexibility that small businesses present.&Support \cite{Walker2007} for small business to develop capabilities around ongoing cyber hygiene and response to breaches.&Government\\
\hline
\rule{0pt}{1ex}~\ref{sec62}&Cohort size - leverage the large number of small businesses.&Further research into feasibility and assistance required in forming networks \cite{Verizon2019,Abolhassan2017} that can be an early warning system within small business cohort. Sharing of knowledge to enable peer support.&Government\\
\hline
\rule{0pt}{1ex}~\ref{sec63}&Zero Trust Model \cite{Kindervag2016} - make use of similarities between zero trust model and common small business architecture.&Further technical research to clarify how the use of zero trust principles can fit into small business' highly hybridised IT environments; Identification of opportunities and development of solutions to leverage overlaps.&Research; Industry\\
\hline
\rule{0pt}{1ex}~\ref{sec64}&Open Source Security Products \cite{Cisco2020,Welte2014,Lyon2019} - repurpose existing low cost solutions to fit small business needs.&Further research into the possibilities and logistics required to leverage open source security products to match small business' operating model and constraints, making use of advantages \cite{Ebart2008,Ven2008,Wesselius2008} that open source offers.&Research\\
\bottomrule
\end{tabular}
\\
\end{minipage}

\pagebreak
Our research found a small business owner's agility and responsiveness could lead to better cyber-security responses.  Rather than focus on formal and process based security standards, a more effective method would be to teach the skills/capabilities a small business needs to both protect itself as well as respond to an incident. Focusing on the skills needed taps into the natural ways in which small businesses excel at solving problems.  

Small businesses could use their large numbers to their advantage by establishing alliances.  Cyber-security alliances are not new and have been shown to benefit members.  To date, alliances have formed between large organisations.  
Small business would obtain definite advantages from such an alliance, especially in terms of peer support, early intelligence and education.

Our research identified a couple of technical cyber-security solutions that could be leveraged for small business cyber-security with the right collaboration and product development.  A zero trust model, by assuming no safe network boundaries, aligns the base IT architecture closer to the small businesses mode of operation.  By discarding a need to ring-fence disparate IT systems, a zero trust model would drive a realistic security architecture plan to protect small businesses. In a similar manner, open source cyber-security software would give small businesses opportunities both in overcoming financial barriers to entry and having their needs met.



\subsection{International Implications}
In this study we detailed the effects of cyber-security decisions/actions by organisations (legal or otherwise) on countries outside their geographical boundaries. Within countries with similar societal profile to Australia, such as UK and US, a similarity in cyber-security has been observed in the evidence produced. The parallels ranged from research approaches e.g. survey, interviews, sole trader/micro business exclusion, to government actions e.g. advisories, frameworks etc. These similarities extended to small business dominance of economy, owners' awareness of cyber-security knowledge gaps, as well as comparable technology skill levels.

Given technology landscapes are also very similar across these countries, it is not inconceivable that small businesses in other countries with similar societal profiles to Australia, such as the UK and US, share similar human struggles with cyber-security.  As such any solutions shown to work in Australia would merit further examinations in other such countries.

The important role of human influence on small business cyber-security leads us to conclude that the small business challenges and characteristics discussed in this paper are not unique to Australian small businesses.

\section{Conclusion}
The urgent need to protect small businesses from cyber-criminals is driven by increasing pressures on small businesses to use technology.  Small businesses face pressure to adopt technology from multiple fronts, ranging from customer expectations to world events. Through our examination of evidence from both research and industry, we have identified gaps in the pursuit of cyber-security for small businesses at multiple levels. 

At a data level, further efforts are needed to clarify and improve understanding around small and micro businesses' ways of working, their IT architecture and real world small business breach loss statistics.  The understanding must be gained through representative small and micro business samples and communicated using standardised security terminologies, so that data and lessons can be transferred in a wider context.

In addition, we discussed small business constraints that impacts on small business's ability to protect themselves.  These range from availability of technical knowledge within the business, organisational maturity and mixed IT ownership. Uncertainties around external factors such as cyber-insurance, legal remediation and cost of cyber-incidents also make small business decisions around cyber-security difficult.

Through our overview approach, opportunities to apply non-traditional solutions to cyber-security are also becoming apparent. Promising characteristics including alliances, a new security paradigm and open source community for helping small businesses to build up their defences were also identified.

With the right research and support, a more coherent understanding of small business cyber-security needs and risks can lead to more resilient small businesses.  It is through a combination of these insights that cyber-security solutions can be made accessible to fit better with small businesses, rather than expect small businesses to fit into current cyber-security solutions. Industry, government and society at large also benefit from the reduced investigative and human costs of cyber-incidents. Supporting small business cyber-security has far reaching benefits, to the economy, the community and national security.

\section{Acknowledgement}
This research is supported by an Australian Government Research Training Program (RTP) Scholarship.

\section{Authors Biographies}
\textbf{Tracy Tam}
is a doctoral candidate (Mathematical Science) within the School of Science at RMIT University. She holds a Bachelor degree (with Honours) in Telecommunications Engineering from Monash University. She worked for a decade as an engineer within the IT industry. She also has experience in managing her own small business. Her research interests include cyber-security within small business context, cyber-security human factors and security risk management.

\textbf{Prof. Asha Rao}
is Professor and Associate Dean (Mathematical Sciences) within the School of Science at RMIT University, Australia. She is an Australian 2019-2020 Superstar of STEM. She has won over AUD 3.5M in grants since 2007 from Government and Industry to research a variety of issues including insider threat and continuous authentication. As a trans-disciplinary researcher, she works on a variety of problems, ranging from designing better codes for communication, exploring the mathematics behind quantum cryptography, finding links between various combinatorial structures, to cybersecurity problems such as risk management for SME, detecting insider threats, and money laundering.

\textbf{Joanne Hall} 
is a Senior Lecturer in mathematics and cybersecurity at RMIT. With a background in abstract algebra, her research is on quantum key distribution and post quantum cryptography.  Dr Hall completed her PhD at RMIT in 2011 on quantum key distribution.  She has held research and teaching positions at Charles University in Prague and the Queensland University of Technology.  As the internships coordinator for the Master of Cybersecurity Degree, Dr Hall has a keen interest in the cybersecurity needs of businesses.

\bibliographystyle{IEEEtranN} 
\bibliography{LitReview}

\end{document}